\documentclass[conf]{new-aiaa}
\usepackage[utf8]{inputenc}
\usepackage{graphicx}
\usepackage{amsmath}
\usepackage{wasysym}
\usepackage{accents}
\usepackage{multirow}
\usepackage{bm}
\usepackage{float}
\usepackage{array}
\usepackage{subfiles}
\usepackage{setspace}
\usepackage[version=4]{mhchem}
\usepackage{longtable,tabularx}
\usepackage{enumitem}
\title{Modeling and Simulation of Virtual Rigid Body Formations and Their Applications Using Multiple Air Vehicles}

\author{Suguru Sato \footnote{Graduate Student, UTA Mechanical and Aerospace Engineering Department, 701 S. Nedderman Drive, Arlington, TX, 76019, AIAA student member.} and Kamesh Subbarao\footnote{Professor, UTA Mechanical and Aerospace Engineering Department, 701 S. Nedderman Drive, Arlington, TX, 76019, and AIAA Associate Fellow}}
\affil{The University of Texas at Arlington, Arlington, Texas, 76019}

\begin{document}

\maketitle

\begin{abstract}
This paper presents thorough mathematical modeling, control law development, and simulation of virtual structure formations which are inspired by the characteristics of rigid bodies. The stable constraint forces that establish the rigidity in the formation are synthesized by utilizing d'Alembert's principle of virtual work, constraint sensitivities (Lagrange multipliers) and constraint stabilization using Baumgarte stabilization. The governing equations of motion of a multiagent system are derived via Newton's and Euler's equations to include these constraint forces and to enable inputs regarding the formation as if it were an independent rigid body. The performance of this framework is evaluated under multiple cases including waypoint following missions, and using different number of agents.
\end{abstract}

\section{Nomenclature}
	{\renewcommand\arraystretch{1.0}
		\noindent\begin{longtable*}{@{}l @{\quad=\quad} l@{}}
		${\bm f}_i $ & force on \(m_i\) \\
		$M_i$ & mass matrix of \(m_i\), \(M_i = m_i I_{3 \times 3}\) \\
		${\bm f}_{i,E}$ & external force on \(m_i\) \\
		${\bm f}_{ij}$ & force on \(m_i\) due to influence from \(m_j\) \\
		${\bm f}_{i,C}$ & constraint force on \(m_i\) \\
		$c_k$ & k\(^{th}\) constraint \\
		${\bm r}_i, {\bm r}_j$ & absolute position of \(m_i\) (or \(m_j\)) \\
		$d_{ij,d}$ & desired distance between \(m_i\) and \(m_j\) \\
		$J$ & jacobian matrix \\
		$\alpha$, $\beta$, $\gamma$ & control law gains \\
		${\bm p}_i$ & linear momentum of \(m_i\) \\
		${\bm \omega}_b$ & angular velocity of the formation with respect to its body frame \\
		$[T]_a^b$ & transformation matrix from \(\{a\}\) frame to \(\{b\}\) frame \\
		${\bm f}_{cm}$ & force on the formation center of mass \\
		${\bm h}_{cm}$ & angular momentum of \({\bm h}_{cm}\) about the formation center of mass \\
		$[I]_{cm}$ & mass moment of inertia of the formation \\
		${\bm f}_{i,ext}$ & external force from surroundings \\
		${\bm f}_{i,u}$ & input force \\
		$\delta(\cdot)$  & relative quantities of \((\cdot)\) \\ 
		$(\cdot)_{a,b}$ & quantity \((\cdot)\) of \(a\) with respect to \(b\) \\
		$(\cdot)^{\times} $ & skew symmetric matrix of quantity \(\cdot\) \\
	\end{longtable*}}

\section{Introduction}
Formation motion, wherein multiple agents operate within close proximity while maintaining precise relative and coordinated movement, has a significant impact in the realm of Uncrewed Aerial Systems (UAS). Advances in this research have led to multiple advantages in areas such as reconnaissance, aerospace exploration, map building, health-care application, search and rescue, etc \cite{Bhatia_2010}. These formation motions and their applications are often inspired by biological systems  \cite{Chung_2009} such as school of fish and flock of birds \cite{Lewis_2014}, as well as natural phenomena such as hydrodynamics (smooth streamline), magnetic field, and electrical charges. In such formulations, formation motion is often captured by a few rules, which are collision avoidance, velocity matching, and flock centering \cite{Lewis_2014}. Moreover, as approaches to satisfy those criteria, there are roughly three methods such as leader-following, behavioral method, and virtual structures \cite{Andre_2005}. Our research is motivated by the characteristics of rigid bodies and can be categorized as virtual structure approach. By considering autonomous agents as fixed nodes on a rigid body, the trajectory of each agent on it is automatically determined based on the behavior and motion of the rigid body. Therefore, there is no need to design their trajectories individually. Moreover, since distances between agents are held constant, collision free trajectory are automatically generated. By borrowing the name from \cite{Zhou_2015}, we call this formation as \emph{Virtual Rigid Body (VRB)} formation since an established formation works as a rigid body while there is no physical connections between nodes/agents. 

The main contributions of this paper are thorough mathematical modeling, control law development, and simulation of the VRB framework for formation motion. The framework synthesizes the 6-DOF equations of motion of a multiagent system for an arbitrary number of agents. Reference motion is prescribed for the aggregated rigid body - velocity of the center of mass and the angular velocity. Using these and the rigid body structure specifications, local trajectories are calculated to perform the formation establishment, reconfiguration, orientation, and station-keeping. Multiple simulations using linear quadratic control methods are performed to illustrate the effectiveness of the modeling framework.

The rest of the paper is organized as follows: The constraint force synthesis for an N-Agents VRB is presented in section III, followed by the equations of motion for the multiagent system in section IV. Section V discusses the development of the simulation setup and results. The summary and conclusions are presented in section VI.

\section{Constraint Force Synthesis for an N-Agent Virtual Rigid Body Formation}
\begin{figure}[H]
\centering
\includegraphics[width = 0.4\textwidth]{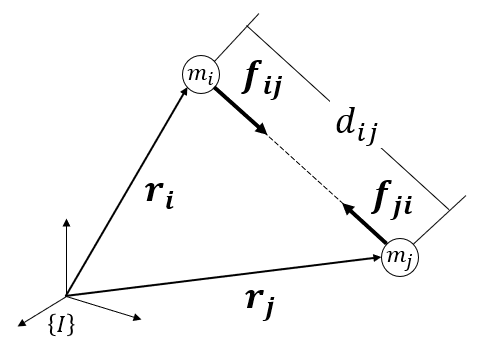}
\caption{Illustration of the relation of \(m_i\) and \(m_j\)}
\label{fig:mij_relation}
\end{figure} 

When a multiparticle system is considered as illustrated in Fig.\ref{fig:mij_relation}, the following describes the translational motion of the system \cite{Schaub_2019}.
\begin{equation}
{\bm f}_i = M_i \ddot{\bm r}_i = {\bm f}_{i,E} + \sum_{j=1, j\neq i}^{N-1} {\bm f}_{ij}
\label{eq:NewtonTranslationalEq}
\end{equation}

The idea of a virtual rigid body formation uses the inter-agent force, \({\bm f}_{ij}\), to constrain the motion of one agent relative to others. Thus denoting \textit{constraint force} as, ${\bm f}_{i,C} = \sum_{j=1, j\neq i}^{N-1} {\bm f}_{ij}$ \cite{Pagilla_2006}, Eq.(\ref{eq:NewtonTranslationalEq}) can be re-written as follows.
\begin{equation}
{\bm f}_i = {\bm f}_{i,E} + {\bm f}_{i,C}
\label{eq:NewtonTranslationalWithConstraint}
\end{equation}

The brief derivation of this constraint force is shown as follows similar to \cite{Pagilla_2006, Bhatia_2011}. From the relation of \(m_i\) and \(m_j\) as well as a desired distance between them, following distance constraint function is developed.
\begin{equation}
c_k = \| {\bm r}_i - {\bm r}_j \| - d_{ij,d}
\label{eq:ConstraintFunction}
\end{equation}

Additionally, for an $N$-particle system in three-dimensional space, the total number of coordinates is 3$N$. The motion of the system as a whole is characterized by 3 translational and 3 rotational motion degrees of freedom (DoF). Since the degrees of freedom are defined as
\begin{equation}
\text{DoF} = \text{number of coordinates} - \text{number of independent constraints},
\end{equation}
the system has $3N-6$ internal degrees of freedom. If these $3N-6$ DoF are fully constrained, the structure becomes rigid. Therefore, $3N-6 (=m)$ independent constraints are required to ensure rigidity.

Moreover, we wish to ensure \(c_k = 0 ~\forall ~k = 1,2,\ldots, m \) in asymptotic manner. To ensure this, the following necessary conditions also need to be satisfied \cite{Pagilla_2006}. 
\begin{equation}
\dot{\bm c} = \frac{d}{dt}{\bm c} = \left[\frac{\partial {\bm c}}{\partial {\bm r}}\right]\frac{d{\bm r}}{dt} = {\bm J}\dot{\bm r} = 0
\label{eq:ConstraintVelocity}
\end{equation}
where, ${\bm J} =  \left[\frac{\partial {\bm c}}{\partial {\bm r}}\right]$
and
\begin{equation}
\ddot{\bm c} = \frac{d}{dt}{\bm J}\dot{\bm r} = \left\{ \frac{d}{dt}{\bm J} \right\}\dot{\bm r} + {\bm J} \left\{ \frac{d}{dt}\dot{\bm r} \right\} = \dot{{\bm J}}\dot{\bm r} + {\bm J}\ddot{\bm r} = 0
\label{eq:ConstraintAcceleration}
\end{equation}

where \({\bm c} = [c_1, c_2, \ldots, c_m]^T\) and \({\bm r} = [r_1, r_2, \ldots, r_N]^T\). From Eq.(\ref{eq:NewtonTranslationalEq}) and (\ref{eq:ConstraintAcceleration}), the expression of the constraint force with the constraint acceleration can be written as follows.
\begin{equation}
{\bm f}_C = \left[ {\bm J}{\bm M}^{-1} \right]^{-1} \left[ -\dot{{\bm J}}\dot{\bm r} - {\bm J}{\bm M}^{-1}{\bm f}_E \right]
\label{eq:ConstraintAccelerationForce}
\end{equation}

Where \({\bm f}_C = [{\bm f}_{1,C}, {\bm f}_{2,C}, \ldots, {\bm f}_{N,C}]^T\), \({\bm f}_E = [{\bm f}_{1,E}, {\bm f}_{2,E}, \ldots, {\bm f}_{N,E}]^T\), and
\begin{equation*}
M = 
\begin{bmatrix}
[M_1] & 0_{3 \times 3} & \ldots & \ldots & 0_{3 \times 3} \\
0_{3 \times 3} & [M_2] & 0_{3 \times 3} & \ldots & 0_{3 \times 3} \\
 & & \vdots & & \\
 0_{3 \times 3} & \ldots & \ldots & 0_{3 \times 3} & [M_N]
\end{bmatrix}
\end{equation*}

This expression of constraint force, Eq.(\ref{eq:ConstraintAccelerationForce}), does not necessarily satisfy Eq.(\ref{eq:ConstraintVelocity}). To ensure Eq.(\ref{eq:ConstraintVelocity}) along with Eq.(\ref{eq:ConstraintAcceleration}), d'Alembert's principle of virtual work, which states work done to the system by constraint force is zero \cite{Schaub_2019}, is applied. 
\begin{equation}
\delta{\bm w}_C = {\bm f}_C \cdot \delta{\bm r} = 0
\label{eq:VirtualWork}
\end{equation}

To ensure this,
\begin{equation}
{\bm f}_C \cdot \delta{\dot{\bm r}} = 0
\label{eq:VirtuaPower}
\end{equation}

From Eq.(\ref{eq:ConstraintVelocity}) and (\ref{eq:VirtuaPower}), following relation can be developed by using the Lagrange Multiplier vector, \(\bm{\lambda}\).
\begin{equation}
{\bm f}_C =  {\bm J}^T \bm{\lambda}
\label{eq:Jlambda}
\end{equation}

By substituting Eq.(\ref{eq:ConstraintAccelerationForce}) into Eq.(\ref{eq:Jlambda}), solving the resultant equation for \({\bm \lambda}\), and substituting the resultant \(\bm{\lambda}\) expression back into Eq.(\ref{eq:Jlambda}), following equation for the constraint force can be obtained.
\begin{equation}
{\bm f}_C =  {\bm J}^T \left[  {\bm J}{\bm M}^{-1} {\bm J}^T \right]^{-1} \left[ - \dot{{\bm J}}\dot{\bm r} -  {\bm J}{\bm M}^{-1}{\bm f}_E \right]
\label{eq:Fc}
\end{equation}

As it appears in the constraint force expression, Eq.(\ref{eq:Fc}), no control law for constraints stabilization is embedded. Therefore if the initial conditions do not fulfill the constraint functions, the solution can diverge over time \cite{Andre_2005}; moreover, choosing initial conditions that satisfies the constraint functions becomes tedious and difficult as the number of agents increases \cite{Pagilla_2006}. Thus, utilization of feedback in the constraint functions must be considered. For the feedback control law to enforce these constraints, Baumgarte stabilization technique \cite{Baumgarte_1971}, is often used for numerical stabilization of multi-body constrained systems \cite{Andre_2005}. By implementing a PID-like control law derived from this Baumgarte stabilization technique, the constraint force is synthesized as follows. Similar work has been done in \cite{Bhatia_2011}.
\begin{equation}
{\bm f}_C =  {\bm J}^T \left[  {\bm J}{\bm M}^{-1} {\bm J}^T\right]^{-1}  \left[ -  {\bm J} {\bm M}^{-1}{\bm f}_E - \dot{ {\bm J}}\dot{\bm r} - \underbrace{2\alpha\dot{\bm c} - \beta^2{\bm c} - \gamma \int_0^t {\bm c}d\tau}_{\text{stabilization term}} \right]
\label{eq:FcFinal}
\end{equation}

The modified constraint force is employed in all formations simulated in this paper.

\section{Equations of Motion of Multiagent System}
In this section, the governing equations of motion of the multiagent system are derived to describe the translational and rotational motion of the VRB, and necessary components such as body frame attachment and input decoupling methods are introduced. Illustration of a multiagent system with a local frame associated with it is shown below.
\begin{figure}[H]
\centering
\includegraphics[width = 0.4\textwidth]{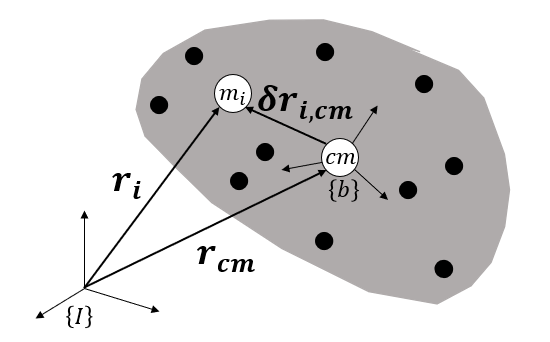}
\caption{Illustraction of a multiagent system}
\label{fig:MultiparticleSystem}
\end{figure}

\subsection*{Translational Dynamics}
The linear momentum of \(M_i\) is expressed as follows.
\begin{equation}
{\bm p}_i = M_i \dot{\bm r}_i
\label{eq:pi_I}
\end{equation}

Since \({\bm r}_i = {\bm r}_{cm} + \delta{\bm r}_{i,cm}\), Eq.(\ref{eq:pi_I}) can be expanded as follows with transport theorem by considering a body frame associated with the system.
\begin{equation}
{\bm p}_i = M_i \left\{ \dot{\bm r}_{cm} + (\delta\dot{\bm r}_{i,cm})_b + {\bm \omega}_b \times (\delta{\bm r}_{i,cm})_b \right\}
\label{eq:pi_b}
\end{equation}

By taking a derivative of Eq.(\ref{eq:pi_b}) with respect to time with employment of transport theorem, following expression for translational dynamics can be obtained.
\begin{equation}
(\delta\ddot{\bm r}_{i,cm})_b = - 2 {\bm \omega}_b \times (\delta\dot{\bm r}_{i,cm})_b - \dot{\bm \omega}_b \times (\delta{\bm r}_{i,cm})_b - {\bm \omega}_b \times \left\{ {\bm \omega} \times (\delta{\bm r}_{i,cm})_b \right\} + [T]_f^b[M_i]^{-1}\left\{ {\bm f}_i - {\bm f}_{cm} \right\}
\label{eq:TranslatinoalDynamicsPre}
\end{equation}

By substituting Eq.(\ref{eq:NewtonTranslationalWithConstraint}) into Eq.(\ref{eq:TranslatinoalDynamicsPre}), equation of translational dynamics is finalized as follows.
\begin{equation}
(\delta\ddot{\bm r}_{i,cm})_b = - 2 {\bm \omega}_b \times (\delta\dot{\bm r}_{i,cm})_b - \dot{\bm \omega}_b \times (\delta{\bm r}_{i,cm})_b - {\bm \omega}_b \times \left\{ {\bm \omega} \times (\delta{\bm r}_{i,cm})_b \right\} + [T]_f^b M_i^{-1}\left\{ ({\bm f}_{i,E} + {\bm f}_{i,C}) - {\bm f}_{cm} \right\}
\label{eq:TranslationalDynamics}
\end{equation}

where \({\bm f}_{cm}\) is the input that is to be calculated and applied on the formation center of mass.

\subsection*{Rotational Dynamics}
Angular momentum of \({\bm h}_{cm}\) about the center of mass is written as follows.
\begin{equation}
{\bm h}_{cm} = \sum_{i=1}^N \delta{\bm r}_{i,cm} \times M_i \delta\dot{\bm r}_{i,cm}
\label{eq:hcm_I}
\end{equation}

By considering that the relative quantities are described in the body frame associated with the multiagent system, Eq.(\ref{eq:hcm_I}) can be expanded.
\begin{equation}
{\bm h}_{cm} = \sum_{i=1}^N (\delta{\bm r}_{i,cm})_b \times M_i \left\{ (\delta\dot{\bm r}_{i,cm})_b + {\bm \omega}_b \times (\delta{\bm r}_{i,cm})_b \right\}
\label{eq:hcm_b}
\end{equation}

Noting that \(\sum_{i=1}^N (\delta{\bm r}_{i,cm})_b^\times (\delta{\bm r}_{i,cm})_b^{\times T} M_i = [I_{cm}]_b\), Eq.(\ref{eq:hcm_b}) is simplified as follows.
\begin{equation}
{\bm h}_{cm} = \sum_{i=1}^N (\delta {\bm r}_{i,cm})_b \times M_i(\delta {\bm r}_{i,cm})_b + [I_{cm}]_b {\bm \omega}_b
\label{eq:hcm_b_simple}
\end{equation}

By taking derivative of Eq.(\ref{eq:hcm_b_simple}) with respect to time and solving the resultant equation for \(\dot{\bm \omega}_b\), following equation of rotational dynamics is obtained.
\begin{equation}
\begin{aligned}
\dot{\bm \omega}_b 
&= -[I_{cm}]_b^{-1} \Biggl[  \left\{{\bm \omega}_b \times \sum_{i=1}^N (\delta{\bm r}_{i,cm}) \right\} \times M_i (\delta\dot{\bm r}_{i,cm})_b + \sum_{i=1}^N (\delta{\bm r}_{i,cm})_b \times M_i (\delta\ddot{\bm r}_{i,cm})_b \\
&+ \sum_{i=1}^N (\delta{\bm r}_{i,cm})_b \times M_i{\bm \omega}_b \times (\delta \dot{\bm r}_{i,cm})_b + [\dot{I}_{cm}]_b {\bm \omega}_b + {\bm \omega}_b \times [I_{cm}]_b {\bm \omega}_b - \sum_{i=1}^N ({\bm r}_{i,cm})_b \times ({\bm f}_i)_b  \Biggl]
\end{aligned}
\label{eq:RotationalDynamics}
\end{equation}

\subsection*{Rotational Kinematics}
To address kinematic singularities resulting from the description using Euler angles and ensure all-attitude capability \cite{Stevens_2016}, Quaternions are utilized for the rotational kinematics.
\begin{equation}
\begin{bmatrix}
\dot{q}_0 \\ \dot{q}_1 \\ \dot{q}_2 \\ \dot{q}_3
\end{bmatrix}
= \frac{1}{2}
\begin{bmatrix}
0 & -p & -q & -r \\
p & 0 & r & -q \\
q & -r & 0 & p \\
r & q & -p & 0
\end{bmatrix}
\begin{bmatrix}
q_0 \\ q_1 \\ q_2 \\ q_3
\end{bmatrix}
\label{eq:RotationalKinematics}
\end{equation}
where, $\bm{q} = \left[ q_0~ q_1~ q_2~ q_3\right]^T$ is the attitude quaternion, and $p$, $q$, and $r$ represent the components of the body angular velocity $\bm{\omega}_b$. Note,
after inegration of Eq.(\ref{eq:RotationalKinematics}), the quaternions are normalized to ensure \(\|{\bm q}\| = 1\).

\subsection*{Translational Kinematics}
Absolute velocity of \(m_i\) can be described as \( \dot{\bm r}_i = \dot{\bm r}_{cm} + \delta\dot{\bm r}_{i,cm}\). Since the relative quantity, \(\delta\dot{\bm r}_{i,cm}\), is described in the local frame associated with the multiagent system, \( \dot{\bm r}_i \) is further expressed as follows from transport theorem.
\begin{equation}
(\dot{\bm r}_i)_I = (\dot{\bm r}_{cm})_I + [T]_b^I \left\{ (\delta\dot{\bm r}_{i,cm})_b + {\bm \omega}_b \times (\delta{\bm r}_{i,cm})_b \right\}
\label{eq:TranslationalKinematics}
\end{equation}

\subsection*{Body Frame Attachment}
To enable simulations using the EOM developed in this section, attachment of local/body frame associated with the multiagent system is required as it appears in the EOM multiple times. Attachment of body frame on a multiagent system is done as a VRB formation is established by utilizing a geometric feature of the formation. For example, formation body frame can be attached as follows for a three-agent case under equilateral triangle formation.
\begin{figure}[H]
\centering
\includegraphics[width = 0.5\textwidth]{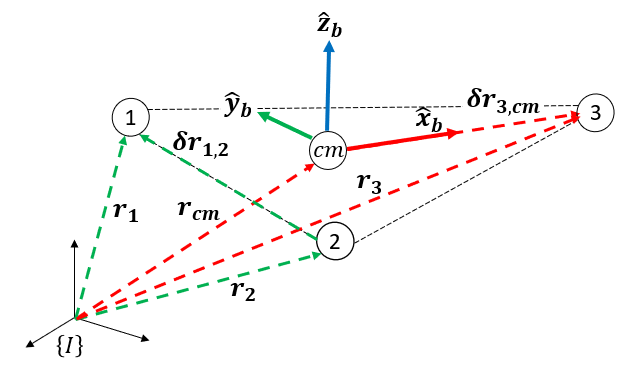}
\caption{Example of body frame attachment}
\label{fig:BodyFrame}
\end{figure}

X-body axis, \(\hat{x}_b\), can be found from the relative position vector between \(m_3\) and the formation center of mass. Y-body axis, \(\hat{y}_b\), is found from a geometric feature that is perpendicular to \(\hat{x}_b\), which is \(\delta{\bm r}_{1,2}\) in Fig.\ref{fig:BodyFrame}. Finally, z-body axis can be found from a cross product of \(\hat{x}_b\) and \(\hat{y}_b\). 
After obtaining body axes at formation establishment, initial intertial-to-body transformation matrix, \([T]_{I,0}^B\), is constructed as such.
\begin{equation}
\begin{bmatrix}
x \\ y \\ z
\end{bmatrix}_b
=
\begin{bmatrix}
C_{11} & C_{12} & C_{13} \\
C_{21} & C_{22} & C_{23} \\
C_{31} & C_{32} & C_{33}
\end{bmatrix}_0
\begin{bmatrix}
x \\ y \\ z
\end{bmatrix}_I
\end{equation}

Where, $[T]_{I,0}^b = \begin{bmatrix}
	C_{11} & C_{12} & C_{13} \\
	C_{21} & C_{22} & C_{23} \\
	C_{31} & C_{32} & C_{33}
\end{bmatrix}_0$. The 
initial quaternion can be calculated from \([T]_{I,0}^B\).

\subsection*{Input Decoupling}
While inputs are calculated under the assumption of the VRB as one rigid body, they need to be distributed to each agent precisely and properly so that the overall motion of VRB due to the motion of each agent represents the desired behavior. 

Since force on center of mass is the summation of all forces on each agent, \({\bm f}_{cm}\) is expressed as follows.
\begin{equation}
({\bm f}_1)_b + ({\bm f}_2)_b + \cdots + ({\bm f}_N)_b = ({\bm f}_N)_b
\label{eq:Fcm}
\end{equation}

Torque on the formation center of mass due to force on \(m_i\) is expressed as follows.
\begin{equation}
({\bm r}_{i,cm})_b^\times ({\bm f}_i)_b = ({\bm \tau}_{cm,i})_b
\label{eq:taucm}
\end{equation}

Equation (\ref{eq:Fcm}) and (\ref{eq:taucm}) can be expressed collectively in a matrix form as follows.

\begin{equation}
\begin{bmatrix}
I_{3 \times 3} & \cdots & \cdots & \cdots & I_{3 \times 3} \\
(\delta{\bm r}_{1,cm})_B^\times & 0_{3 \times 3} & \cdots & \cdots & 0_{3 \times 3} \\
0_{3 \times 3} & (\delta{\bm r}_{2,cm})_B^\times & 0_{3 \times 3} & \cdots & 0_{3 \times 3} \\
\vdots & \vdots & \vdots & \vdots & \vdots \\
0_{3 \times 3} & \cdots & \cdots &\cdots & (\delta{\bm r}_{N,cm})_B^\times
\end{bmatrix}
\begin{bmatrix}
({\bm f}_{1})_B\\
({\bm f}_{2})_B\\
\vdots \\
\vdots \\
({\bm f}_{N})_B
\end{bmatrix}
=
\begin{bmatrix}
({\bm f}_{cm})_B \\ 
({\bm \tau}_{cm,1})_B \\
({\bm \tau}_{cm,2})_B \\
\vdots \\
({\bm \tau}_{cm,N})_B
\end{bmatrix}
\label{eq:Fcmtaucm}
\end{equation}

By abbreviating the matrix in Eq.({\ref{eq:Fcmtaucm}}) as \({\bm H}\), input on each agent is calculated as follows.
\begin{equation}
\begin{bmatrix}
({\bm f}_{1})_b \\
({\bm f}_{2})_b \\
\vdots \\
\vdots \\
({\bm f}_{N})_b
\end{bmatrix}
=
({\bm H}^T{\bm H})^{-1}{\bm H}^T
\begin{bmatrix}
({\bm f}_{cm})_b \\ 
({\bm \tau}_{cm,1})_b \\
({\bm \tau}_{cm,2})_b \\
\vdots \\
({\bm \tau}_{cm,N})_b
\end{bmatrix}
\label{eq:pinv}
\end{equation}

\section{Simulations}
In this section, developments from the previous sections are verified via multiple simulations. Once VRB establishment from the constraint force, Eq.(\ref{eq:FcFinal}), is confirmed, components from section IV are synthesized, then simulated to finalize the viability of the framework. Note, the individual agents all implement a local linear quadratic regulator (LQR) based control laws to track their desired trajectories. The desired local trajectories are trivially determined from the overall rigid body motion specification. The motion (translation and rotation) of the VRB is converted to $N$ agent trajectories (from the rigid body specification).

\subsection*{Formation Establishment}
Simulation architecture of formation establishment by the constraint force is show below.
Following values are provided as inputs.
\begin{table}[H]
	\centering
	\begin{tabular}{ |p{3cm}||p{12cm}| }
		\hline
		\multicolumn{2}{|c|}{\textbf{Input specifications}} \\
		\hline
		\textbf{Input Description} & \textbf{Values} \\
		\hline
		Number of Agents & \(N = 3\) \\
		Agent's mass   & \(m_i = 1 kg \: \forall \: i = 1, 2, 3\)    \\
		Desired distances & \([d_{12,d}, d_{13,d}, d_{23,d}] = [4, 4, 4]m\:[0, 6]s, [4, 8, 4]m\:[6, 13]s, [4, 4, 4]m\:[13,20]s\) \\
		Initial position & \({\bm r}_{1,0} = [1, 6, 3]^Tm, {\bm r}_{2,0} = [8, 3, 3]^Tm, {\bm r}_{3,0} = [7, 6, 3]^Tm\) \\
		Initial velocity & \({\bm v}_{i,0} = [0,0,0]^Tm/s \: \forall \: i = 1, 2, 3\) \\
		External force &  \({\bm f}_{i,ext} = [0,0,-m_ig]^TN\) (gravitational force) \\
		Input force & \({\bm f}_{i,u} = [0.1,0.1,m_ig]^TN\) (constant altitude translation) \\
		\hline
	\end{tabular}
	\caption{Formation establishment input specifications}
	\label{tab:table1}
\end{table}

\begin{figure}[H]
\centering
\includegraphics[width = 0.6\textwidth]{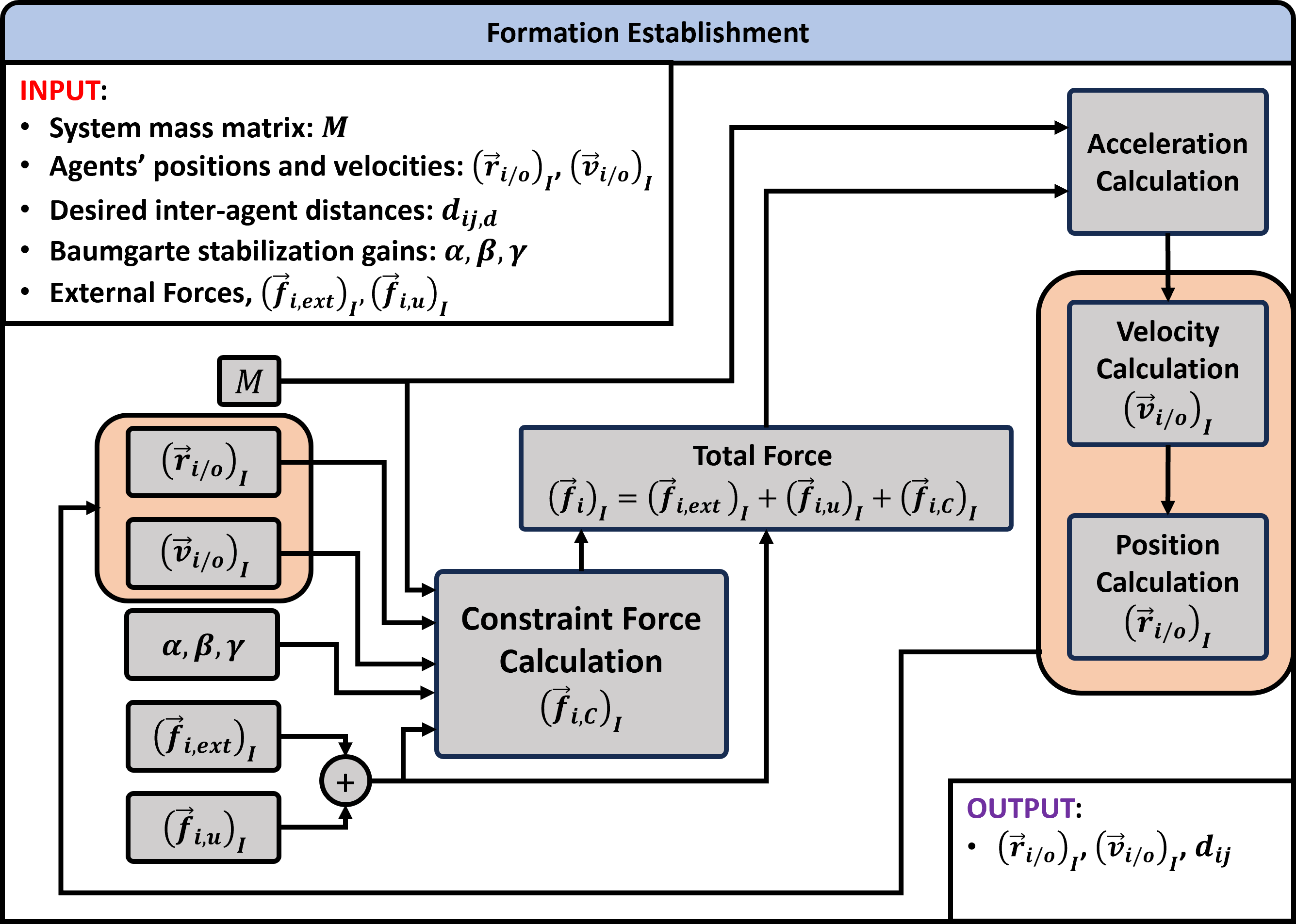}
\caption{Formation establishment simulation architecture}
\label{fig:FormationEstablishmentArchitecture}
\end{figure}

By Implementing this architecture in MATLAB, following results for formation establishment was obtained.
\begin{figure}[H]
\centering
\includegraphics[width = 0.9\textwidth]{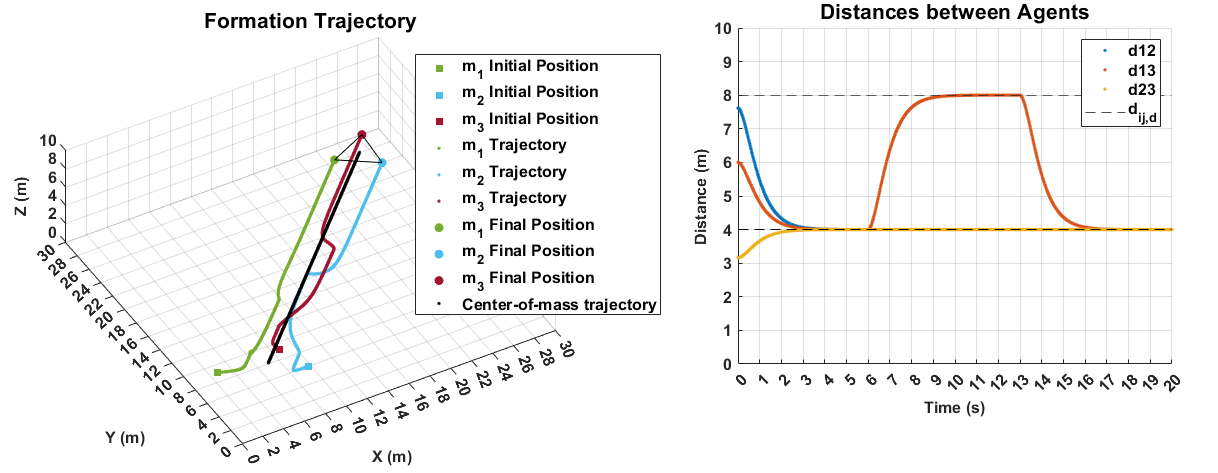}
\caption{Formation establishment with constant altitude translational motion}
\label{fig:FormationEst_Combined}
\end{figure}

As Fig.\ref{fig:FormationEst_Combined} shows, formation establishment at an equilateral triangle formation as well as formation reconfiguraitons to a linear formation based on distance constraints are satisfied and stabilized at the desired inter-agent distances while the multiagent system translates. Thus, formation establishment and formation reconfiguration are validated. Formation establishment for other cases with different number of agents are shown in the Appendix.

\subsection*{Multiagent EOM Synthesis and Single Waypoint Simulation}
Since formation establishment was confirmed, the multiagent EOM and the necessary components developed in the previous section are integrated and simulated with control inputs calculated from LQR \cite{Lewis_2011} aimed a single waypoint flight. Simulation architecture of this work is shown below.

\begin{figure}[H]
\centering
\includegraphics[width = 0.7\textwidth]{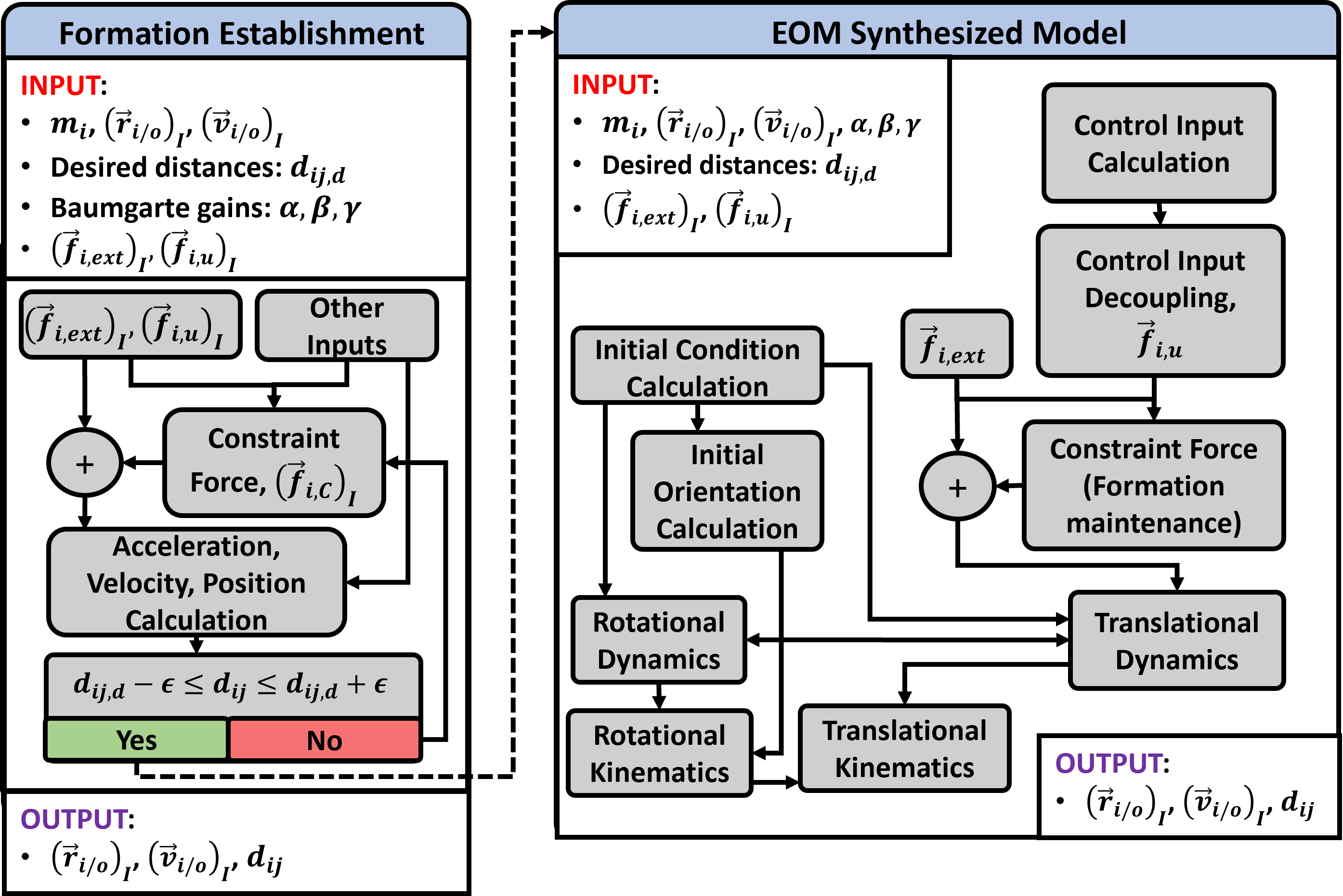}
\caption{Simulation architecture of the framework with EOM implemented}
\label{fig:EOM_synthesis}
\end{figure}

Simulation inputs as well as the predefined waypoint specifications are shown in the table below.

\begin{table}[H]
\centering
\begin{tabular}{ |p{4cm}||p{9cm}| }
 \hline
 \multicolumn{2}{|c|}{\textbf{Simulation Input and Waypoint Specifications}} \\
 \hline
 \textbf{Simulation Input} & \textbf{Values} \\
 \hline
 Number of Agents & \(N = 3\) \\
 Agent's mass   & \(m_i = 1 kg \: \forall \: i = 1, 2, 3\)    \\
 Desired distances & \(d_{ij} = 4m \: \forall \: ij = 12, 13, 23\) \\
 Initial position & \({\bm r}_{1,0} = [1, 6, 3]^Tm, {\bm r}_{2,0} = [8, 3, 3]^Tm, {\bm r}_{3,0} = [7, 6, 3]^Tm\) \\
 Initial velocity & \({\bm v}_{i,0} = [0,0,0]^Tm/s \: \forall \: i = 1, 2, 3\) \\
 External force &  \({\bm f}_{i,ext} = [0,0,-m_ig]^TN\) (gravitational force) \\
 \hline
 \textbf{Waypoint} & \textbf{Values} \\
 \hline 
 C.M. position & \({\bm r}_{cm} = [15, 15, 15]^Tm\) \\
 C.M. velocity &  \({\bm v}_{cm} = [0, 0, 0]^Tm/s\)\\
 Formation orientation & \({\bm \sigma} = [0, 0, -90]^{\circ T}\) \\
 Formation angular rates & \({\bm \omega}_b = [0, 0, 0]^{\circ T}/s\)  \\
 \hline
 \textbf{Method} & \textbf{Approach} \\
 \hline
 Linear-Quadratic-Regulator & Hamiltonian \\
 min. state and input energy & \(J = \frac{1}{2}X^T(T)S(T)X(T) + \frac{1}{2}\int_0^T(X^TQX + U^TRU)dt\) \\
 \hline
\end{tabular}
 \caption{Simulation specifications for single waypoint case}
 \label{tab:table2}
\end{table}

By integrating the required components in Simulink, following simulation results are obtained.
\begin{figure}[H]
\centering
\includegraphics[width = 0.9\textwidth]{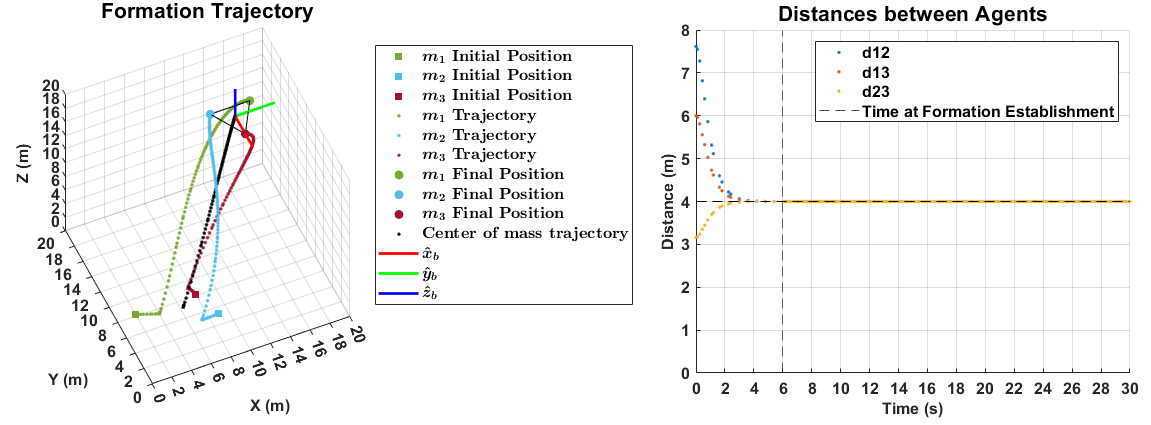}
\caption{Single waypoint simulation formation trajectory and \(d_{ij}\) results}
\label{fig:SWP_Combined}
\end{figure}

\begin{figure}[H]
\centering
\includegraphics[width = 0.9\textwidth]{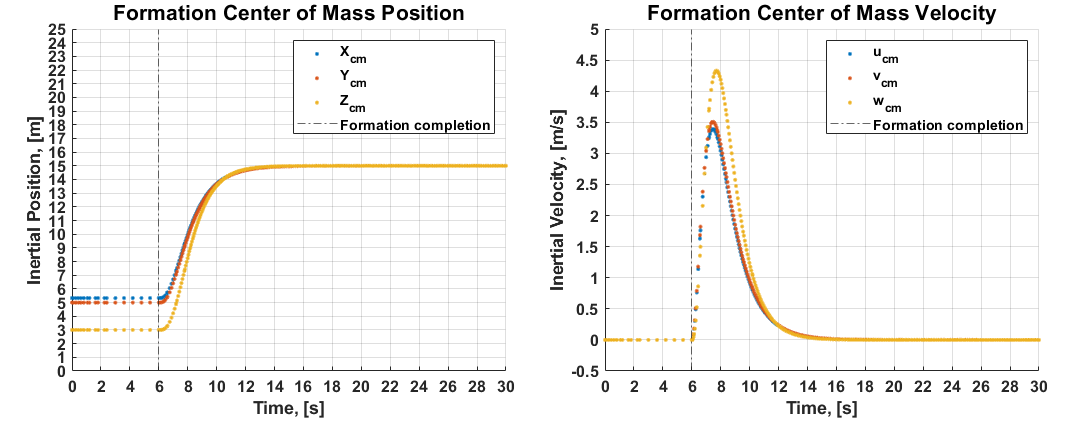}
\caption{Single waypoint simulation C.M. position and velocity results}
\label{fig:SWP_RcmVcmCombined}
\end{figure}

\begin{figure}[H]
\centering
\includegraphics[width = 0.9\textwidth]{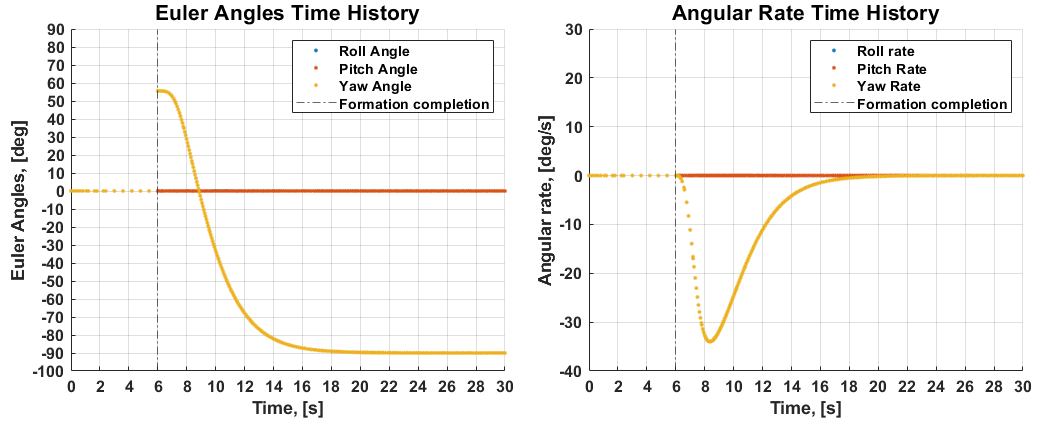}
\caption{Single waypoint simulation formation Euler angle and angular velocity results}
\label{fig:SWP_OrientationCombined}
\end{figure}

As Fig.\ref{fig:SWP_Combined} shows, the formation from desired distances between agents are established and stabilized for the rest of the simulation while it is maneuvering to certain states. Moreover, as Fig.\ref{fig:SWP_RcmVcmCombined} shows, application of control inputs begins after the formation is achieved, and desired position and velocity of the formation center of mass is achieved and stabilized. Finally, as Fig.\ref{fig:SWP_OrientationCombined} shows, once formation is achieved, initial orientation of the formation is calculated, and desired orientation as well as desired angular velocity are realized and stabilized for the rest of the simulation period.

\begin{figure}[H]
\centering
\includegraphics[width = 0.6\textwidth]{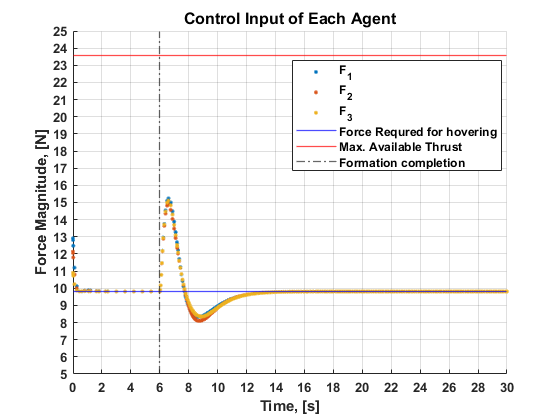}
\caption{LQR single waypoint simulation, input magnitude for each agent}
\label{fig:inputs}
\end{figure}
According to \cite{Liu_2023}, the ASL (Aerospace System Laboratory) quadcopter model approximately has maximum available thrust of 5.9 N for each rotor including thrust
required for hovering. Therefore, the total available thrust for the quadcopter is
approximately 23.6 N. As Fig.\ref{fig:inputs} shows, the control input applied to each agent is
all well below the maximum available thrust. Since our agents are currently point
masses, thrust required for each rotor is unknown. Therefore, when implementation
of this framework into the physical system is considered, further investigation needs
to be carried out.

\subsection*{Multiple Waypoint Simulation}
The framework from single waypoint simulation is brought into a multiple waypoint simulation in an artificially developed environment. Following environment as well as key scenarios are considered to check the viability of the framework under multiple occasions.
\begin{figure}[H]
\centering
\includegraphics[width = 1.0\textwidth]{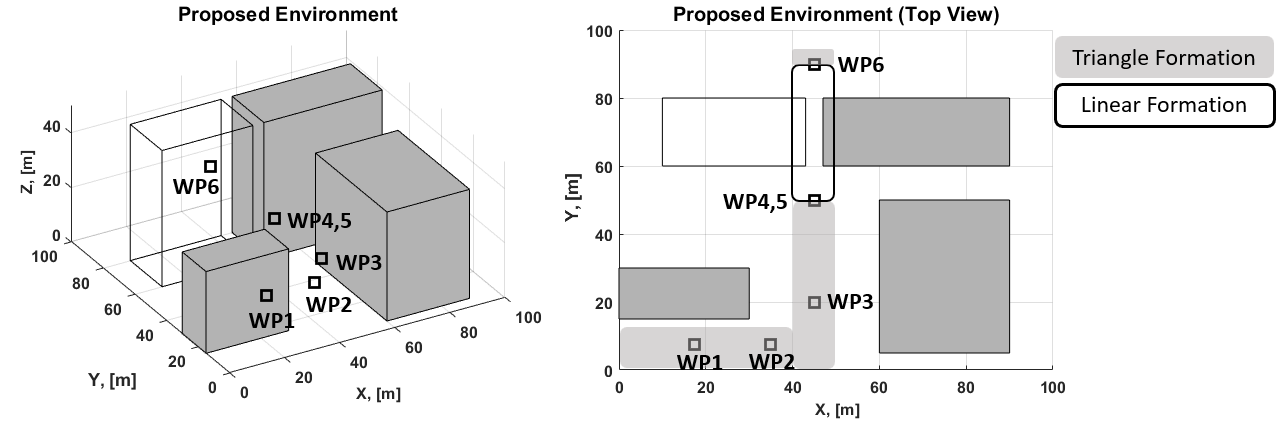}
\caption{Illustration of simulation environment}
\label{fig:Environment_Combined}
\end{figure}

\begin{table}[H]
\centering
\begin{tabular}{ |p{2cm}||p{13cm}| }
 \hline
 \multicolumn{2}{|c|}{\textbf{Waypoints Specifications and Descriptions}} \\
 \hline
 \textbf{Waypoint} & \textbf{Specifications} \\
 \hline
 Waypoint 1 & \({\bm r}_{cm} = [17.5, 7.5, 20]^Tm, {\bm v}_{cm} = [0,0,0]^Tm/s, {\bm \sigma} = [0,0,0]^{T\circ}, {\bm \omega}_b = [0,0,0]^{T\circ}/s\) \\
 Waypoint 2 & \({\bm r}_{cm} = [35, 7.5, 20]^Tm, {\bm v}_{cm} = [2,0,0]^Tm/s, {\bm \sigma} = [0,0,0]^{T\circ}, {\bm \omega}_b = [0,0,0]^{T\circ}/s\) \\ 
 Waypoint 3 & \({\bm r}_{cm} = [45, 20, 20]^Tm, {\bm v}_{cm} = [0,2,0]^Tm/s, {\bm \sigma} = [0,0,0]^{T\circ}, {\bm \omega}_b = [0,0,0]^{T\circ}/s\) \\ 
 Waypoint 4 & \({\bm r}_{cm} = [45, 50, 20]^Tm, {\bm v}_{cm} = [0,0,0]^Tm/s, {\bm \sigma} = [0,0,0]^{T\circ}, {\bm \omega}_b = [0,0,0]^{T\circ}/s\) \\ 
 Waypoint 5 & \({\bm r}_{cm} = [45, 50, 20]^Tm, {\bm v}_{cm} = [0,0,0]^Tm/s, {\bm \sigma} = [0,0,90]^{T\circ}, {\bm \omega}_b = [0,0,0]^{T\circ}/s\) \\
 Waypoint 6 & \({\bm r}_{cm} = [45, 90, 20]^Tm, {\bm v}_{cm} = [0,0,0]^Tm/s, {\bm \sigma} = [0,0,90]^{T\circ}, {\bm \omega}_b = [0,0,0]^{T\circ}/s\) \\
 \hline
 \textbf{Key Phases} & \textbf{Descriptions} \\
 \hline 
 Waypoint 0 & Clear the ground \\
 Waypoint 1 & Reach certain altitude to gain safe altitude \\
 Waypoint 4 & Reconfigure the formation from equilateral triangle to linear \\
 Waypoint 5 & Re-orient the formation to the narrow path ahead  \\
 Waypoint 6 & Retrieve the original equilateral triangle \\
 \hline
\end{tabular}
 \caption{Simulation specifications for multiple waypoint case}
 \label{tab:table3}
\end{table}

Following simulation architecture is considered.
\begin{figure}[H]
\centering
\includegraphics[width = 1.0\textwidth]{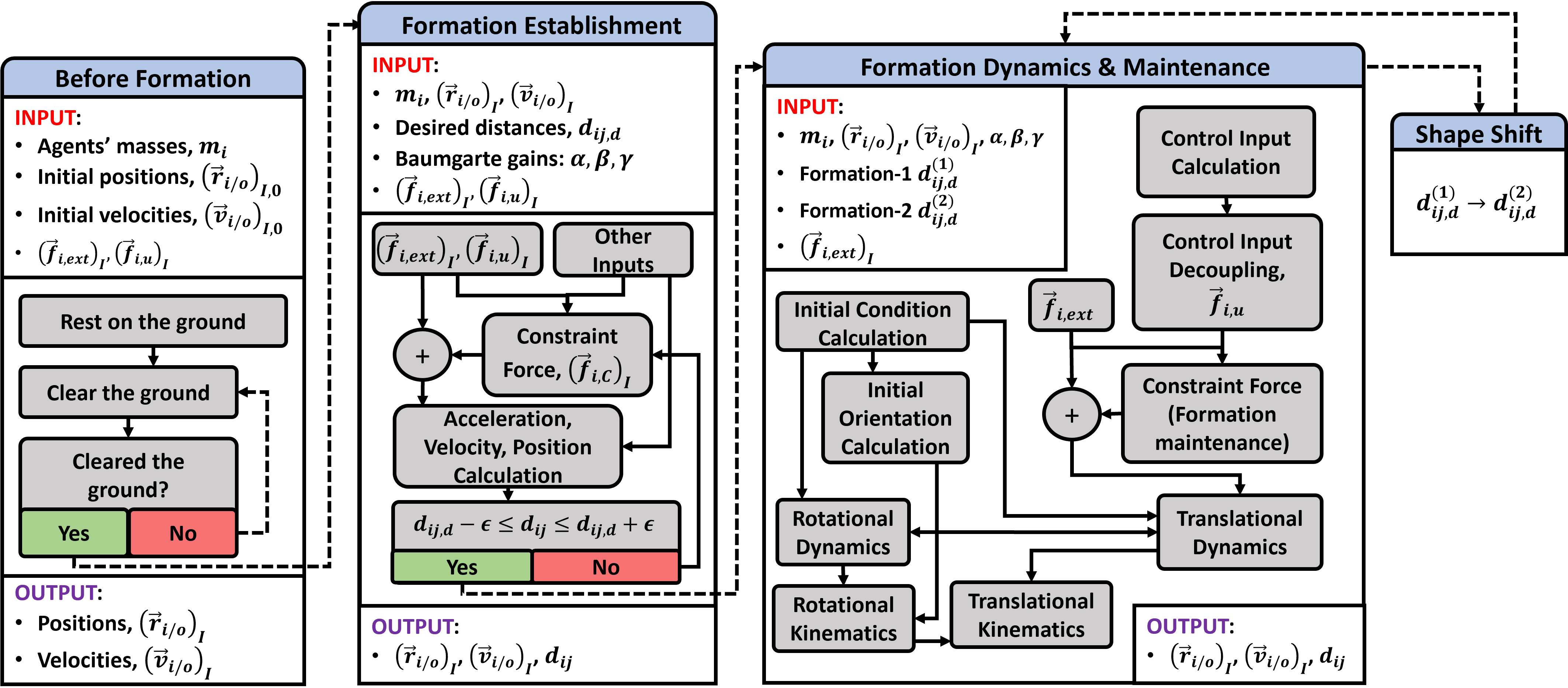}
\caption{Multiple waypoint - reconfiguration simulation architecture}
\label{fig:EnvironmentSim_Architecture}
\end{figure}

As a result of the simulation, following results are obtained.
\begin{figure}[H]
\centering
\includegraphics[width = 0.9\textwidth]{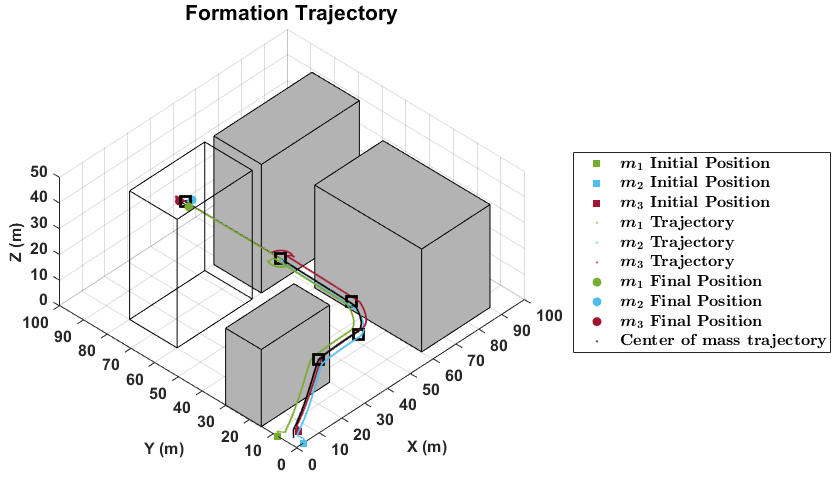}
\caption{Multiple waypoint simulation formation trajectory result}
\label{fig:EnvironmentSim_Trajectory}
\end{figure}
\begin{figure}[H]
\centering
\includegraphics[width = 0.5\textwidth]{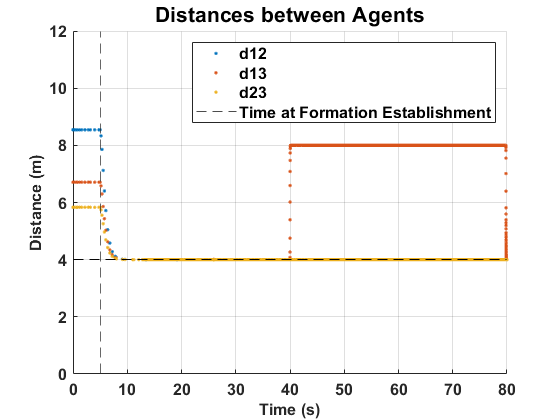}
\caption{Multiple waypoint simulation inter-agent distance result}
\label{fig:EnvironmentSim_Distance}
\end{figure}
\begin{figure}[H]
\centering
\includegraphics[width = 0.9\textwidth]{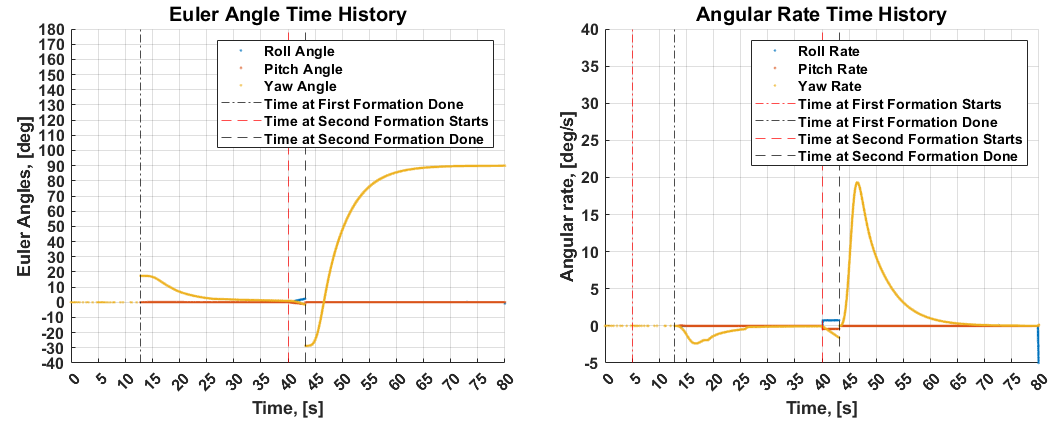}
\caption{Multiple waypoint simulation formation Euler angle and angular velocity results}
\label{fig:EnvironmentSim_EulerRates_Combined}
\end{figure}

As Fig.\ref{fig:EnvironmentSim_Trajectory} and \ref{fig:EnvironmentSim_Distance} show, multiple-waypoint flight are successful. More importantly, formation reconfiguration to the linear formation is realized before the narrow path, and formation retrieve to triangle formation is also achieved followed by successful clearance of the narrow path. As Fig.\ref{fig:EnvironmentSim_EulerRates_Combined} shows, desired formation orientation as well as angular velocity for both before and after the formation reconfiguration are achieved. Deviation in both Euler angle and angular rates are observed upon formation reconfiguration since there exists a change in mass moment of inertia in such a situation.
In this section, the scale of the framework is increased from three agents to eight agents to confirm the viability of the framework with higher number of agents. Beginning from formation establishment, single waypoint and multiple waypoint flight are performed.

\subsection*{Formation Establishment}
The same simulation architecture as Fig.\ref{fig:FormationEstablishmentArchitecture} is applied for this simulation. Following distance constraints and input specifications are considered.
\begin{figure}[H]
	\centering
	\includegraphics[width = 0.2\textwidth]{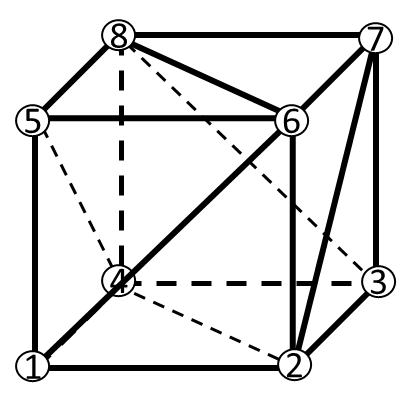}
	\caption{Eight-agent case constraint graph}
	\label{fig:CubeConstraintGraph}
\end{figure}
As \cite{Pagilla_2006} states, the maximum number of independent constraint for 3D formation is \(3N-6\), and any number of constraints over this value will result in an over-defied system. Therefore, eighteen constraints which is shown in Fig.\ref{fig:CubeConstraintGraph} are carefully chosen for this eight-agent case.
\begin{table}[H]
	\centering
	\begin{tabular}{ |p{4cm}||p{11cm}| }
		\hline
		\multicolumn{2}{|c|}{\textbf{Input specifications}} \\
		\hline
		\textbf{Input Description} & \textbf{Values} \\
		\hline
		Number of Agents & \(N = 8\) \\
		Agent's mass   & \(m_i = 1 kg \: \forall \: i = 1, 2, \ldots, 8\)    \\
		Desired distances, \(d_{1j,d}\) & \([d_{12,d}, d_{14,d}, d_{15,d}, d_{16,d}] = [4, 4, 4, 4\sqrt{2}]m\) \\
		Desired distances, \(d_{2j,d}\) & \([d_{23,d}, d_{24,d}, d_{26,d}, d_{27,d}] = [4, 4\sqrt{2}, 4, 4\sqrt{2}]m\) \\
		Desired distances, \(d_{3j,d}\) & \([d_{34,d}, d_{37,d}, d_{38,d}] = [4, 4, 4\sqrt{2}]m\) \\
		Desired distances, \(d_{4j,d}\) & \([d_{45,d}, d_{48,d}] = [4, 4\sqrt{2}]m\) \\
		Desired distances, \(d_{5j,d}\) & \([d_{56,d}, d_{58,d}] = [4, 4]m\) \\
		Desired distances, \(d_{6j,d}\) & \([d_{67,d}, d_{68,d}] = [4, 4\sqrt{2}]m\) \\
		Desired distances, \(d_{7j,d}\) & \(d_{78,d} = 4m\) \\
		Initial position & \({\bm r}_{1,0} = [2, 1, 1]^Tm, {\bm r}_{2,0} = [3, 1, 1]^Tm, {\bm r}_{3,0} = [5, 2, 1]^Tm, {\bm r}_{4,0} = [1, 5, 1]^Tm,\) 
		\({\bm r}_{5,0} = [4, 5, 4]^Tm, {\bm r}_{6,0} = [5, 4, 4]^Tm, {\bm r}_{7,0} = [7, 8, 4]^Tm, {\bm r}_{8,0} = [5, 8, 4]^Tm\) \\
		Initial velocity & \({\bm v}_{i,0} = [0,0,0]^Tm/s \: \forall \: i = 1, 2, \ldots, 8\) \\
		External force &  \({\bm f}_{i,ext} = [0,0,-m_ig]^TN\) (gravitational force) \\
		Input force & \({\bm f}_{i,u} = [0.1,0.1,m_ig]^TN\) (constant altitude translation) \\
		\hline
	\end{tabular}
	\caption{Eight-agent case formation establishment input specifications}
	\label{tab:tableEightAgent1}
\end{table}

As a result of simulation, following formation establishment outcomes are obtained.
\begin{figure}[H]
	\centering
	\includegraphics[width = 1.0\textwidth]{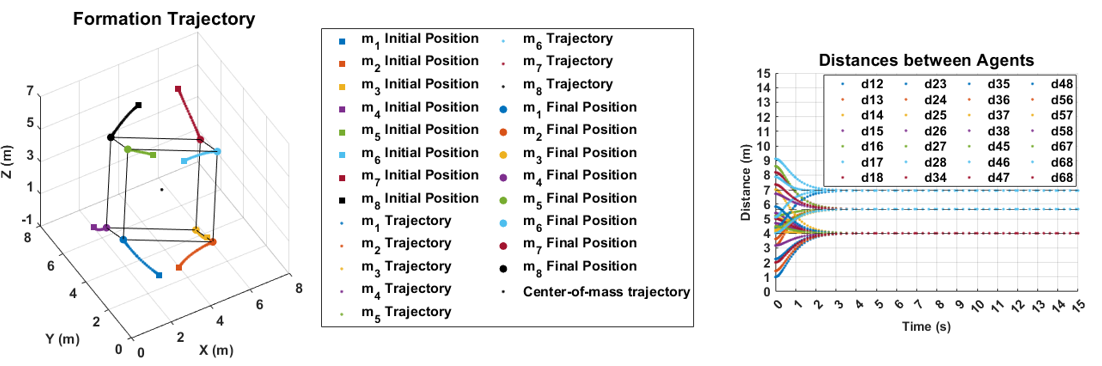}
	\caption{Cube formation establishment results}
	\label{fig:Cube_Combined}
\end{figure}

As Fig.\ref{fig:Cube_Combined} shows, distance constraints based on the desired distances are achieved and stabilized, and the expected formation configuration is realized.

\subsection*{Single Waypoint Simulation}
Since formation establishment is confirmed, body frame is attached to the formation as follows, and single waypoint flight simulation utilizing LQR is performed. Specifications of the waypoint flight is shown followed by the body frame attachment step.
\begin{figure}[H]
	\centering
	\includegraphics[width = 0.2\textwidth]{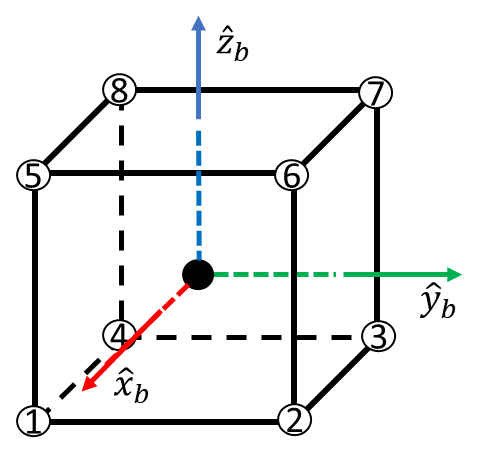}
	\caption{Cube formation body frame attached}
	\label{fig:CubeBodyFrame}
\end{figure}
\begin{table}[H]
	\centering
	\begin{tabular}{ |p{4cm}||p{9cm}| }
		\hline
		\multicolumn{2}{|c|}{\textbf{Waypoint Specifications}} \\
		\hline
		\textbf{Waypoint} & \textbf{Values} \\
		\hline 
		C.M. position & \({\bm r}_{cm} = [10, 4.25, 4.5]^Tm\) \\
		C.M. velocity &  \({\bm v}_{cm} = [0, 0, 0]^Tm/s\)\\
		Formation orientation & \({\bm \sigma} = [0, 0, 0]^{\circ T}\) \\
		Formation angular rates & \({\bm \omega}_b = [0, 0, 0]^{\circ T}/s\)  \\
		\hline
	\end{tabular}
	\caption{Waypoint specificaitons for single waypoint case}
	\label{tab:tableCubeSWP}
\end{table}

For the input specifications, the same values as Table \ref{tab:tableEightAgent1} are applied, and the same input calculation method and approach from Table \ref{tab:table2} are also applied. As a result, following simulation results are obtained.

\begin{figure}[H]
	\centering
	\includegraphics[width = 1.0\textwidth]{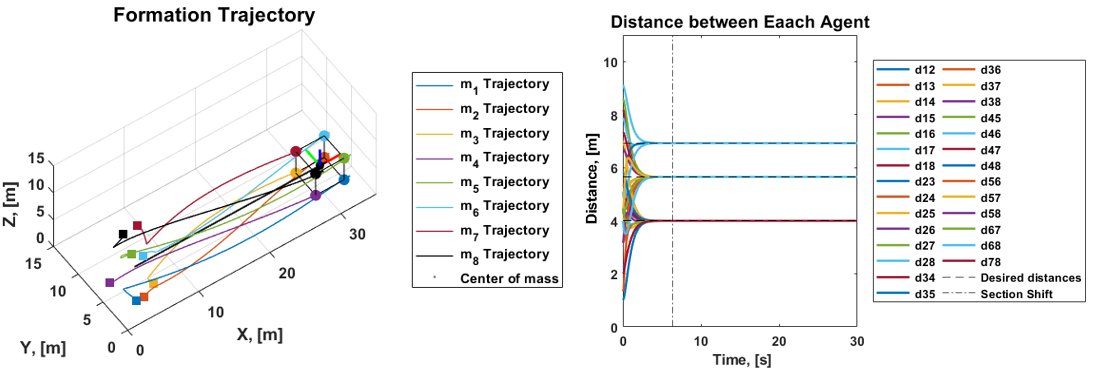}
	\caption{Eight-agent case single waypoint formation trajectory and inter-agent distance results}
	\label{fig:Eight_SWP_Combined}
\end{figure}
\begin{figure}[H]
	\centering
	\includegraphics[width = 1.0\textwidth]{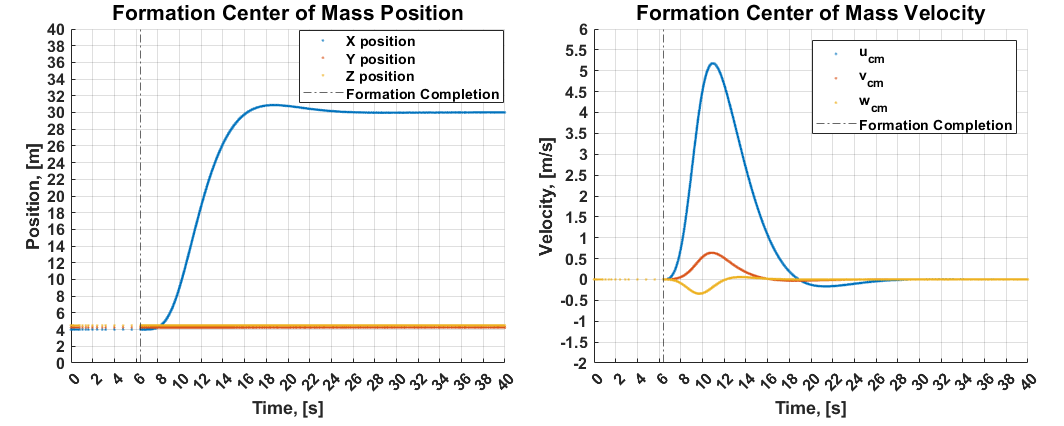}
	\caption{Eight-agent formation center of mass position and velocity results}
	\label{fig:Eight_SWP_RcmVcm}
\end{figure}
\begin{figure}[H]
	\centering
	\includegraphics[width = 1.0\textwidth]{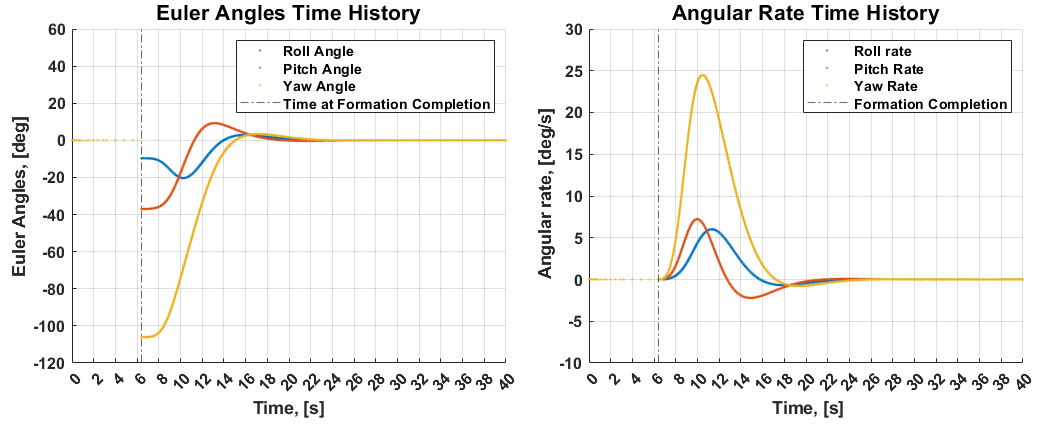}
	\caption{Eight-agent formation Euler angle and angular velocity results}
	\label{fig:Eight_SWP_EulerRates}
\end{figure}

As Fig.\ref{fig:Eight_SWP_Combined} shows, all the distance constraints are satisfied and stabilized for the entire simulation time period, and the cubic formation derived from them are also realized. As Fig.\ref{fig:Eight_SWP_RcmVcm} shows, both of formation center of mass position and velocity reached their desired values and stabilized. Finally, as Fig.\ref{fig:Eight_SWP_EulerRates} shows, the orientation of the formation is initialized as it is established, and both formation Euler angles and angular velocities are stabilized at the desired values. From these results, it can be concluded that this framework is viable for the higher scale of formations.

\subsection*{Multiple Waypoint Simulation}
Since validation of the framework under single waypoint situation is confirmed, the framework is brought into the environment consists of multiple waypoint situation. The same waypoint specifications are applied; however, at waypoint 4, formation is reconfigured from the cube to a cuboid formation, and the formation is rotated at the destination instead of formation retrieve to show the rotational capability under multiple waypoint situation.
\begin{figure}[H]
	\centering
	\includegraphics[width = 1.0\textwidth]{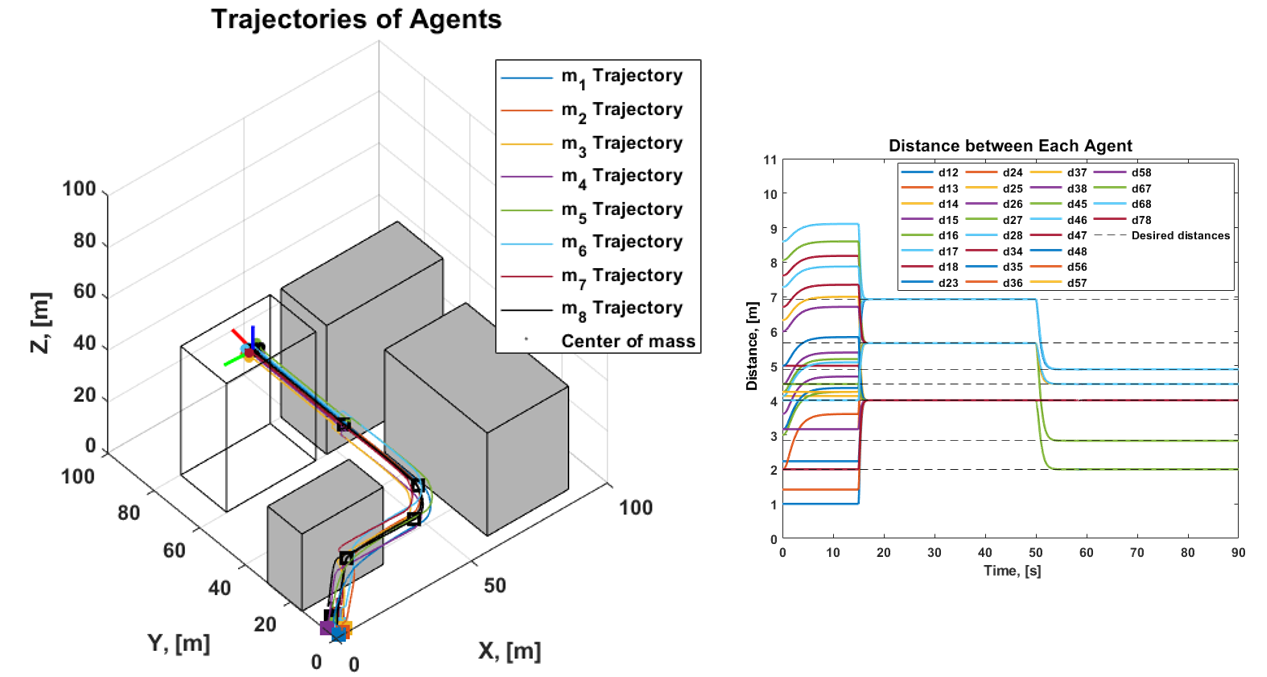}
	\caption{Eight-agent case multiple waypoint simulation formation trajectory and inter-agent distance results}
	\label{fig:Eight_MWP_Combined}
\end{figure}
\begin{figure}[H]
	\centering
	\includegraphics[width = 1.0\textwidth]{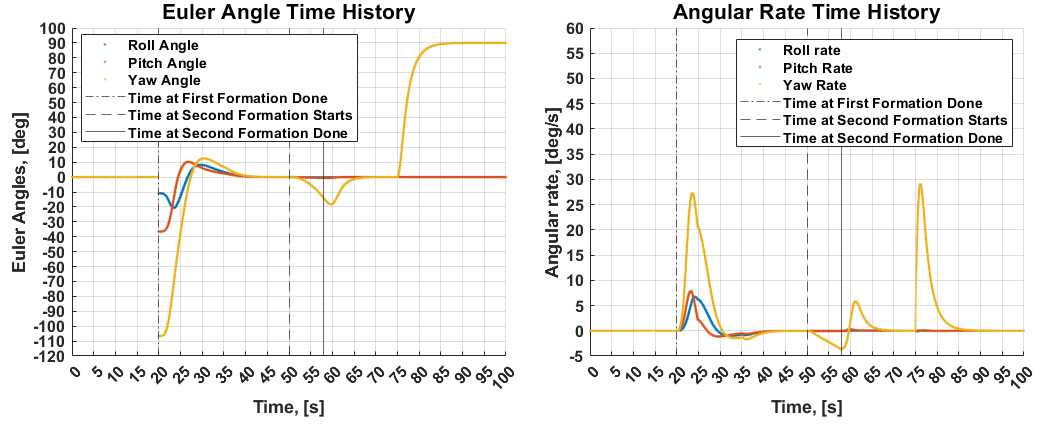}
	\caption{Eight-agent case multiple waypoint simulation, formation Euler angles and angular velocity results}
	\label{fig:Eight_MWP_EulerRates}
\end{figure}

As Fig.\ref{fig:Eight_MWP_Combined} shows, distance constraints for cubic formation is satisfied after all agents cleared the ground around 15 seconds in the simulation. Around 50 seconds in the simulation, reconfiguration to the cuboid simulation started, and the corresponding distance constraints are satisfied shortly after. As Fig.\ref{fig:Eight_MWP_EulerRates} shows, Euler angles of the formation is initialized for the first time after the cubic formation is established, and both orientations and angular velocities reached and were stabilized at the desired values before the formation reconfiguration. As reconfiguration of the formation to cuboid formation began around 50 seconds in the simulation, deviations in rotational properties occurred; however, they were corrected once the cuboid formation was established. Around 75 seconds in the simulation, final rotation of the formation at the destination began, and the formation orientation and angular velocity reached and was stabilized at their desired values. According to these results, it can be concluded that this framework is also capable of performing multiple waypoint missions using an input calculation algorithm with higher number of agents, which widens the potential of this framework to ranges of applications.

\section{Summary and Conclusion}
In this paper, we have developed the stable constraint force algorithm which contains a PID-like Baumgarte stabilization technique motivated by the characteristics of a rigid body. This force and the control algorithm act to satisfy the distance constraints. Equations of motion of a multiagent system are also derived to embed the constraint force and to enable inputs about the center of mass and torque input about the formation body frame. Within the rotational kinematics equations, quaternions are utilized to remove kinematic singularities due to other common attitude parameterizations such as the Euler angles. Along with the equations of motion, input decoupling method as well as body frame attachment method are introduced to complete the formation framework. We have shown that formation reconfiguration is possible with desired inter-agent distances. Once viability of the constraint is confirmed, all components were integrated and tested via simulations with inputs calculated through an LQR based feedback control law. In the single waypoint simulation, desired values for all states were realized and stabilized for different number of agents. Similarly from the multiple waypoint simulations it was concluded that formation establishment, formation reconfiguration, formation reorientation, and station keeping are all possible for different number of agents. Future work will address the collision avoidance among agents as a part of this framework.

\newpage
\section*{Appendix}
In this appendix, simulations for formations with different number of agents as well as different formation configurations are shown.
\begin{figure}[H]
\centering
\includegraphics[width = 0.8\textwidth]{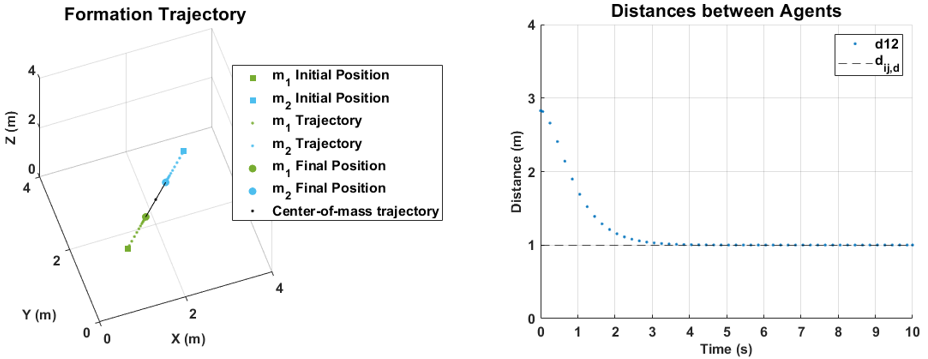}
\caption{Two-agent case, formation establishment results}
\label{fig:TwoPM_Combined}
\end{figure}
\begin{figure}[H]
\centering
\includegraphics[width = 0.8\textwidth]{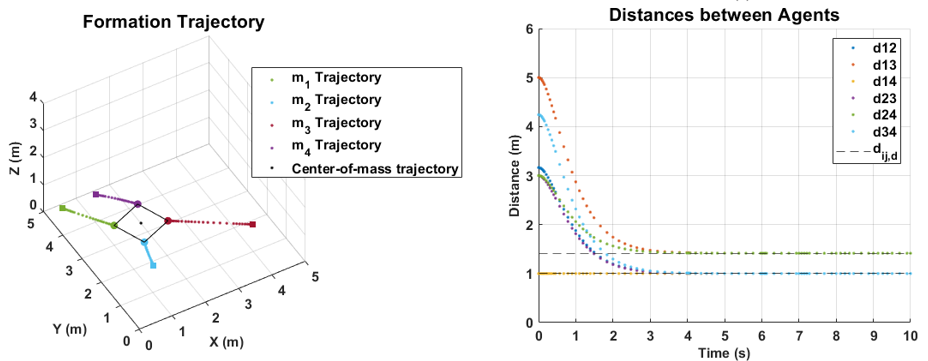}
\caption{Four-agent case, square formation establishment results}
\label{fig:FourPM_Combined}
\end{figure}
\begin{figure}[H]
\centering
\includegraphics[width = 0.8\textwidth]{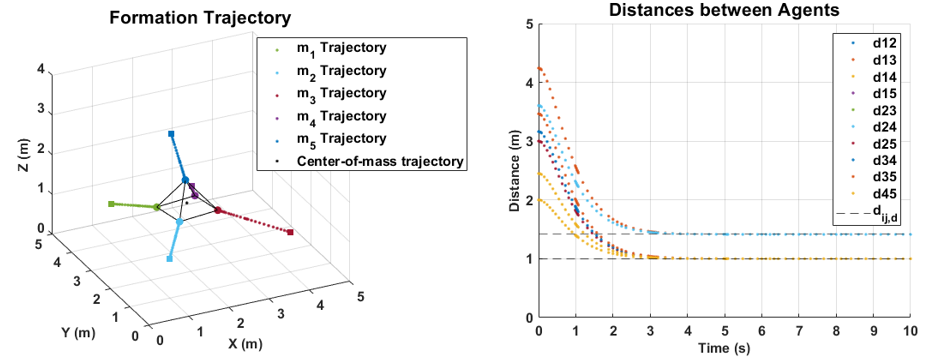}
\caption{Five-agent case, pyramid formation establishment results}
\label{fig:FivePM_Combined}
\end{figure}
\begin{figure}[H]
\centering
\includegraphics[width = 0.8\textwidth]{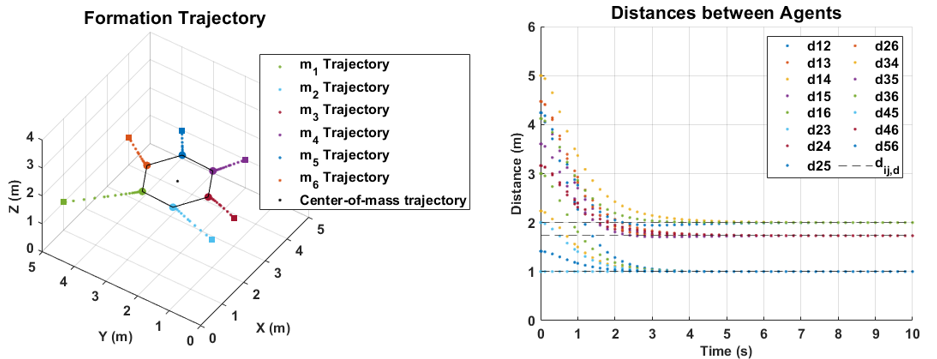}
\caption{Six-agent case, hexagon formation establishment results}
\label{fig:SixPM_Combined}
\end{figure}

As Fig.\ref{fig:TwoPM_Combined} - \ref{fig:SixPM_Combined} shows, desired distances are reached and stabilized for all cases, and desired formations derived from those desired inter-agent distances are also realized. Besides the shown formations configurations, various formation shapes can be established by applying desired inter-agent distances obtained from desired formation configurations as input variables.

\newpage

\bibliographystyle{plain} 

\end{document}